# Investigation of Monolayer-formation time for the Synthesis of Graphene on copper/nickel/silicon using very-low Pressure Pulses of Methane

V-M. Freire[*], A. Ramírez, E. Pascual, E. Bertran, and J-L. Andújar

*FEMAN Group, IN²UB, Departament de Física Aplicada i Òptica, Universitat de Barcelona*, Martí i Franquès 1, E08028 Barcelona, Spain.

**ABSTRACT.** *Chemical Vapor Deposition (CVD) of graphene on copper is one of the most efficient technologies for producing high quality graphene for large areas. Nevertheless, still high pressures and big quantities of precursor gas are currently required. The objective of this work is to deposit graphene using the monolayer-formation-time concept ($\tau$) from the kinetic theory of gases, which leads to an economization of the precursor gas, a minimization of the process pressure and the time needed to grow a graphene monolayer. Our process has been designed taking into account the dependence of $\tau$ on the pressure, the mass particle of the gas, the sticking coefficient, and the growth temperature. Thus, with this alternative method, based on CVD but using very-low pressure instant pulses of precursor gas (~$10^{-4}$ Pa), we have reduced the deposition time to the order of 10 s. We carried out the processes at temperatures below 1000 ºC with methane ($CH_4$) as a precursor gas under High Vacuum (HV) conditions. After Raman spectroscopy and mapping, and Scanning Electron Microscopy (SEM) characterization of the samples, the results point to the formation of high quality monolayer graphene on sputtered copper and silicon substrates covering domain areas of $10^4$ $\mu m^2$.*

---

[*] Corresponding author. Tel.: +34 934039224; fax: +34 934039219.
  E-mail address: victor.freire@ub.edu (V-M. Freire).

## I. INTRODUCTION

The present work is devoted to the growth of graphene[1], the very well-known monatomic layer composed of carbon atoms densely packed in a hexagonal lattice. This structure is also the basis of other allotropes of carbon such as graphite, carbon nanotubes and fullerenes[2]. An extensive review of the properties can be found in references[3-5]. Those extreme properties have made the number of graphene-related papers increase exponentially in the last 8 years[6] possibly because the potential applications of graphene seem to be huge[7]. However, because of the high cost involved in the manufacturing of high quality large domain graphene sheets, there are still limitations to its industrial implementation.

CVD of graphene on copper has been reported as an efficient technology for producing high quality graphene sheets for large areas. Although it meets both requirements in quality and cost, still high pressures and big flows of the precursor gases must be used[8,9]. The segregation/precipitation of carbon atoms from the bulk metal[10] during the annealing and cooling stages, gives place to the graphite formation. But incidentally, the nucleation on Cu is a self-limited process due to the low solubility of carbon in Cu; it starts in some specific sites and the process theoretically stops when the surface is fully covered[11,12]. In principle we could expect it as a self-limited process, but in the practice it is not so clear[13-15]. There is still much to do in the understanding of the growth mechanisms.

The aim of this work is to study the growth of monolayer graphene with a very low-pressure Pulsed-CVD technique by means of a theoretical approach, the monolayer formation time, which has never been used before for this purpose.

## II. EXPERIMENTAL

### 2.1 Experimental equipment

All the processes involved in this work were performed in a custom reactor with a spherical main chamber and having two different pre-chambers. In one of them, a quartz tube for CVD processing is mounted inside a tubular furnace (Fig. 1).

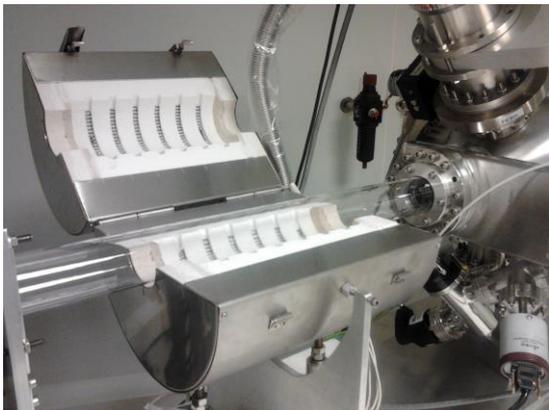

**Figure 1.** View of a small part of the reactor of the UB Clean Room (GRAPHMAN) used in this work. The CVD chamber, the quartz tube, and the cylindrical oven are shown in detail.

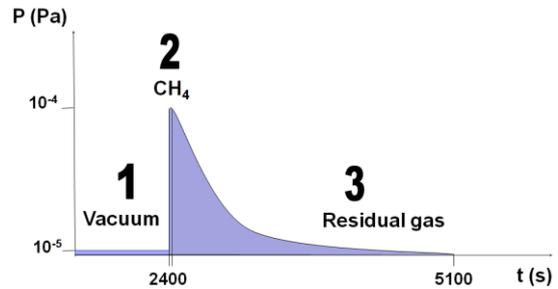

**Figure 2.** P(t) diagram of the whole Pulsed-CVD process. In the step 1) the reactor is under HV conditions while a linear ramp temperature up to 1000 ºC is applied during 40 min, the step 2) corresponds to the instant release of a $CH_4$ pulse of $10^{-4}$ Pa during 10 s, and the final step 3) only with the residual gas and the quenching ramp to room temperature during 45 min.

The other pre-chamber is equipped with a magnetron sputtering system (600 W of nominal RF power, water cooled and 3" targets). In addition, the reactor has a load-lock chamber system to avoid atmospheric contamination into the main chamber during the substrate introduction and among the different CVD steps. The vacuum system is composed by a double stage mechanical pump for the primary vacuum and a turbomolecular pump for achieving high vacuum. The base pressure achieved of the vacuum chamber is $\sim 10^{-5}$ Pa. An automatic butterfly valve of variable conductance, located between the chamber and the turbomolecular pump, keeps constant the pressure conditions during the growing processes.

One of the novelties is the gas introduction system. A sequence of 4 independent valves connected in series (2+2) and a depressurization chamber in the middle of them; and all together connected to a main bottle of precursor gas. This system can deliver the precursor gas by means of very-low-pressure pulses under control. Gas pulses, total pressure, and plasma process were computer-controlled with a LabView® interface.

### 2.2 Substrates

Cu films on c-Si wafers were deposited by magnetron sputtering. We chose sputtered Cu because sputtering allows obtaining pristine and very smooth Cu films with desired thicknesses. Cu has been commonly used as a catalyst for the graphene growth in the Cu foil shape. Nevertheless, not many literature of sputtered Cu has been published[16]. A previous calibration of the Cu deposition rate was performed by profilometry (DEKTAK). Cu films of 600 nm were deposited on 2" diameter and 300 $\mu$m thick (111) c-Si wafers. Native $SiO_2$ layer of the c-Si wafer can play an important role in limiting the diffusion of the Cu in Si when the sample is annealed[17]. However, a Ni diffusion barrier turned out to be much more effective, because after annealing tests, samples with a Ni barrier of 100 nm succesfully avoided a complete diffusion between Cu and Si.

Furthermore, the crystallographic orientation of Cu, preferably (111), is of fundamental importance in the quality and domain size of graphene. A mismatch at the edges of the graphene precludes obtaining large area graphene domains[18].

*2.3 Graphene growth process*

Normally, the most relevant parameters in a standard CVD process are: pressure, flow of the reactant gases, substrate temperature and surface substrate. The experimental conditions used are based on previous publications [8,12,13,19]. However, in this work we used a very low-pressure CVD method based on the monolayer formation time concept. From the kinetic theory of gases one can calculate the time required to form a monomolecular or monoatomic layer on a gas-free surface. This monolayer formation time is closely related with the so-called impingement rate $\Gamma$.

With a gas at rest the impingement rate $\Gamma$ (or particle flux) will indicate the number of molecules which collide with the surface inside the vacuum vessel per unit of time and surface area: $P$ is the pressure, $m$ is the mass of the particle, $k$ is the Boltzmann's constant and $T$ is the temperature, from ideal gas law[20,21]:

$$\Gamma = \frac{P}{(2\pi mkT)^{1/2}} \quad (1)$$

The inverse of the gas impingement rate is related to the monolayer formation (or coverage) time. If $S$ is the sticking coefficient, $a$ is the number of spaces per unit of surface area (a graphene surface has ~$3.6 \times 10^{19}$ sites/m$^2$); then the monolayer formation time is, in SI units[21]:

$$\tau = \frac{a}{S\Gamma} = a\sqrt{(2\pi k)}\frac{\sqrt{mT}}{SP} = 3.4 \times 10^8 \frac{\sqrt{mT}}{SP} \quad (2)$$

On the assumption that every molecule will stick to the surface (sticking coefficient, $S = 1$), working temperatures of 1273 K (1000 ºC), and using methane ($CH_4$) as a precursor gas, one-atom thick layer of carbon spread onto a surface can be formed in a time defined by:

$$\tau = \frac{2.1 \times 10^{-3}}{P} \quad (3)$$

What it means that a monolayer formation time could only be dependent on the gas pressure of the precursor gas. Thus, for a ~$2 \times 10^{-3}$ Pa of methane, the monolayer formation time would be ~1 s.

Once the pressure and the process time were fixed, we placed the substrates in the CVD quartz tube of the reactor, which was pumped down during approximately 30 min until a base pressure of ~$10^{-5}$ Pa was achieved.

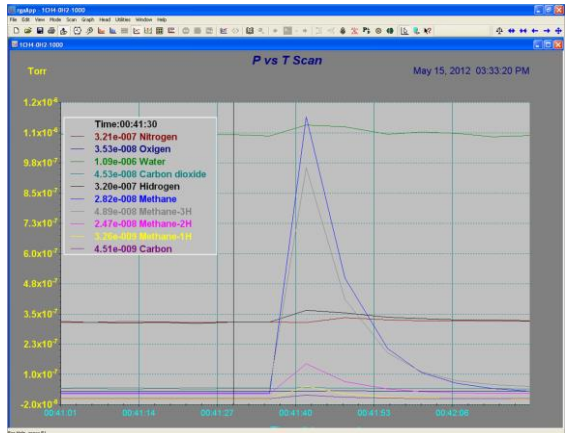

**Figure 3.** Screen capture of the QMS controller, *rgaApp*. The methane pulse was monitorized and its decomposition in the different radicals due to the temperature: $CH_3$, $CH_2$, $CH$, and $C$; also the presence of other common gases as $H_2$, $N_2$, $H_2O$, $O_2$… The pressure of the methane pulse is ~$10^{-3}$ mTorr (~$10^{-4}$ Pa).

**Table I.** A series of 5 samples regarding the substrates (Cu and Si) and the CVD process temperatures is shown. Note that we needed to apply more pulses in sample #4 due to the presence of $H_2$ (two pulses of $10^{-4}$ Pa). The corresponding Raman spectra and the laser intensity are plotted in Fig. 4.

| Sample | Temperature (ºC) | Substrate | # Layers | # Pulses of $CH_4$ ($10^{-4}$ Pa) |
|---|---|---|---|---|
| 1 | 980 | Cu | 1 | 1 |
| 2 | 980 | Si | 1 | 1 |
| 3 | 990 | Cu | 2 | 1 |
| 4 | 990 | Si | 1 | 4 (with $H_2$) |
| 5 | 1000 | Cu | 1 | 1 |

The first step in the CVD process consisted in a linear temperature ramp of the cylindrical oven with a heating rate of 25 ºC/min up to the process temperature reaches 970-1000 ºC. Methane decomposes at temperatures above 800 K,[22] so the methane CVD processes must be performed at a minimum temperature of 527 ºC.

The CVD process started once the substrate reached the process temperature, the methane was then introduced in the chamber under a controlled pulse of ~$10^{-4}$ Pa of pressure. Under these conditions, surface temperature of the substrate induces the decomposition of the methane molecules in different radical species ($CH_4$, $CH_3$, $CH_2$, $CH$, and $C$), which become adsorbed, whereas the gas pumping removes the rest of by-products. During the process, $H_2$ can be introduced to catalyse the reaction (reducing the copper oxide present on Cu, not very suitable as a catalyst) and to drag the by-products (radicals). However, the reductive action of hydrogen exposure at high temperature is known to efficiently etch graphene and to limit its growth. A hydrogen atmosphere is also expected to suppress carbon enrichment at high temperature at the defect sites in Cu, thus to allow to efficiently inhibit carbon segregation[14]. Finally, the CVD process ended with a cooling ramp switching off the oven until room temperature. Fig. 2 shows a P(t) diagram of the entire process. The complete CVD process and the methane pulse was monitorized by a Residual Gas Analyzer (SRS RGA300), a quadrupolar mass spectrometer (QMS) attached to the reactor, and it allows to measure simultaneously the partial pressures of a maximum of 10 gases inside the chamber by means of the relation $m/q$. The screen capture of the QMS controller in Fig. 3 shows some of the radical species involved in the process among of other typical compounds inside the reactor: $N_2$, $H_2O$, $O_2$, $H_2$...

In past works acetylene was used[13,15], a less common and poorly explored precursor of graphene growth[8,19], with CVD temperatures of 800-900 ºC. Acetylene pyrolysis starts at lower temperature than methane pyrolysis. Thus, at a given temperature, a higher deposition rate is expected using acetylene instead of methane[23]. However, the lower deposition rate of methane leads to longer process times, which helps carbon atoms to nucleate and self-assembly on the substrate surface and, therefore, to increase the quality of graphene, which continues to be dependent on temperature[8].

We have observed that quality of graphene depends on the quenching curve too. Fast-cooling processes have been used to suppress the amount of precipitated carbon. However, this process still yields films with a wide range of graphene layer thicknesses, from one to a few tens of layers and with defects associated with fast cooling[24]. Medium cooling gives graphene, and slow cooling has nothing on the surface in that carbon atoms diffuse deep into the bulk catalyst.

Therefore, we opened the oven cover when the CVD chamber achieved a temperature of 800 °C. From that point, the quenching time reduced to half an hour. Several samples were produced by means of the above described process. See the details of some of them in Table I.

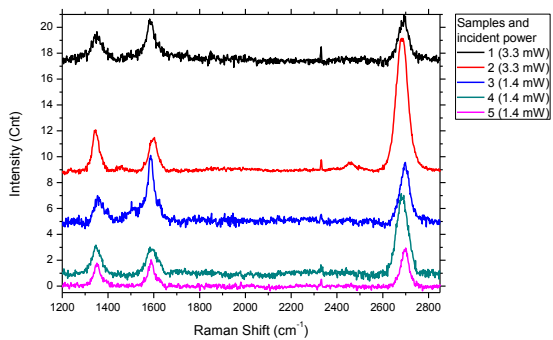

**Figure 4.** Collection of Raman spectra of the Table I samples including the incident power. The acquisition time for all the spectra is always 30 s, and the 2D/G ratio is always ≥1, which confirms the presence of mono-bi layer graphene. However, the presence of defects is somehow constant probably because of the grain formation in the annealed sputtered surfaces.

### 2.4 Characterization

Scanning electron microscopy (SEM) images have been acquired with a Hitachi S4100 microscope with a field emission gun. Also, qualitative measurements of chemical composition were performed by thermionic emission Jeol JSM 840 and Cambridge S120 microscopes, both equipped with an energy-dispersive X-ray spectroscopy (EDS) system.

Raman spectroscopy is the most used technique for a fast graphene identification and characterization. Three main peaks can be observed, namely G, 2D, and D bands.

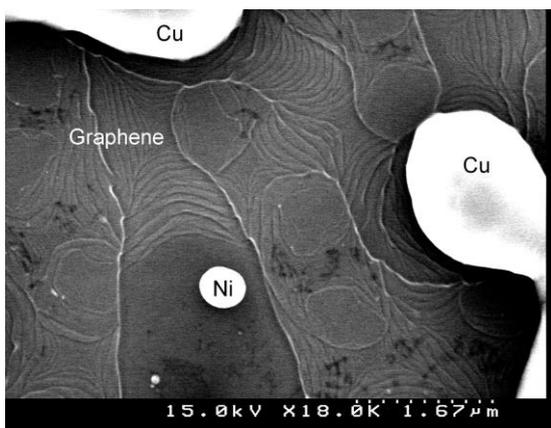

**Figure 5.** SEM image of a graphene sample. This image shows the dewetting of the Cu/Ni bilayer: a Ni island (bright), Cu (bright) and the graphene terraces growth on Si (dark). This process usually takes place during the annealing of the Cu/Ni bilayer previously sputtered on c-Si. The annealing of the bilayer was performed during the CVD under a temperature of 980 °C and an annealing time of 7 min.

The G band appears at ~1580 cm$^{-1}$, its intensity increases with the number of graphene layers. The 2D band, at ~2700 cm$^{-1}$, is the second order effect of the D band and it does not depend on the presence of defects.

The D band only appears in presence of defects or in the edges of a graphene flake, it is used as a measure of quality (located at ~1350 cm$^{-1}$). The characterization relies to this 2D band because its shape and its intensity ratio with G band strongly depend on the number of graphene layers[25]. The contrast can still be improved adjusting the irradiation wavelength and changing the type of substrate[26,27]. We acquired the Raman spectra with a Jobin Yvon LabRam HR 800. The excitation wavelength used was 532 nm from a solid state laser. This green light was applied and collected with a 100X objective under micro-Raman conditions, where the diameter of the analyzed spot is about 1 $\mu$m. The details of the power and the acquisition time that we normally used are given in the spectra (Fig. 4). On the other hand, Raman mapping was performed by a Witec Confocal Raman Microscope Alpha 300R.

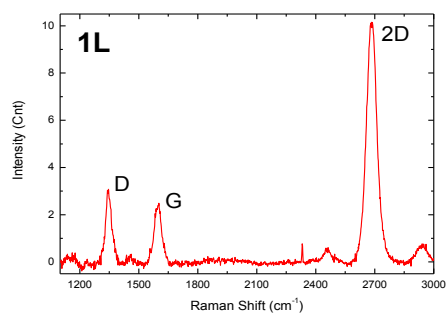

**Figure 6.** Raman spectrum of a graphene monolayer of sample 2 (Si). It was acquired with a 532 nm laser, 3.3 mW of power and an acquisition time of 30 s. The 2D peak is approximately four times the G peak, which corresponds to monolayer graphene[25]. Still a small amount of defects can be observed due to non-flat surface.

### III. RESULTS AND DISCUSSION

Graphene was produced by CVD on Cu as described before. The study of the Cu/Ni and graphene layers deposition consisted on a morphological surface characterization by optical, confocal, and SEM microscopy; and a structural and chemical surface analysis by Raman spectroscopy and EDS. SEM image of Fig. 5 shows the effect of the 7 min annealing time carried out at 980 °C. The continuous phase (bright color) corresponds to Cu crystals (from 1-5 $\mu$m), the dark zones correspond to Si substrate, and Ni (bright color) appears forming clusters separated inside the Si zones and completely separated from the Cu phase. This image shows the dewetting isolating a Ni island and the graphene growth on Si. Also, how the graphene wrinkles formed on graphene terraces overlap. This process usually takes place during annealing of the Cu/Ni bilayer previously sputtered on c-Si.

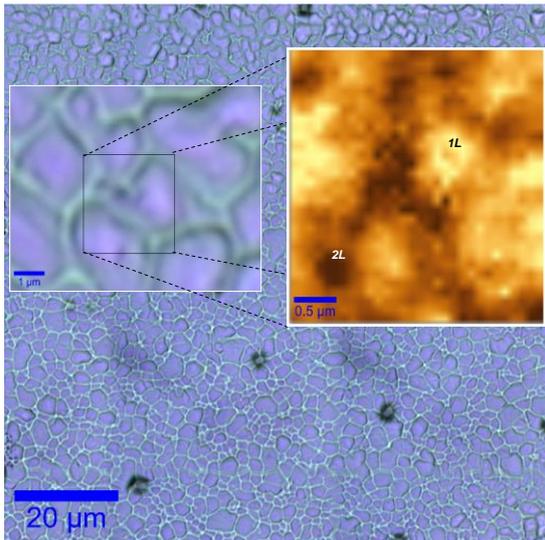

**Figure 7.** Composition of the optical image of the sample with the corresponding Raman mapping acquired with the Witec Raman microscope. The intensity ratio between 2D and G peaks is depicted with >1 (monolayer graphene (yellow)), and ~1 (bilayer graphene (dark brown)). Most of the surface (~80%) is covered by monolayer graphene (1L, yellow) and the rest by bilayer graphene (2L, dark brown).

In previous works[13], a thin Cu film was deposited on a silicon wafer with no native oxide. Because of the film thickness, it was expected that diffusion induced by a thermal treatment could strongly modify the surface of the Cu. EDS indicated the formation of a particular copper silicide compound[28,29] $Cu_{3.17}Si$.[13] However, the complete dewetting process can be avoided by increasing the thickness of the Cu films and introducing a Ni diffusion barrier. On the other hand, if we use very thin Cu layers, dewetting and evaporation of the Cu could be controlled during or immediately after the CVD process looking for the direct deposition of carbon on the silicon substrate[28]. This could be interesting to grow graphene directly on silicon without the need to transfer it. Successful growth of continuous films of graphene with that Cu thickness and type of substrate were reported[11,12].

Graphene appeared continuous and covered almost the entire surface with domains of few tens of $\mu$m in size. In some edges of these domains the graphene layers apparently overlap one on the other (Fig. 5). This is consistent with the idea that carbon layers start to nucleate in some specific sites and then expand during the CVD process until fully covering the Cu surface[12].

The graphene grown on Cu has normally 2-3 layers of graphene, while graphene grown on Si is always mono or bilayer. This is because, first graphene grows on Cu and later, upon evaporation during annealing of Cu, graphene is transferred onto bare Si areas. In this process, the carbon could be rearranged, reducing G peak, which appeared as a reduction in the number of layers; and reducing D peak, a reduction in the defects.

Several research papers[10,29-33] have reported that wrinkles like those shown in our SEM images are folds of graphene and the slightly darker areas are additional graphene layers. Hence, the graphene grown by this method still has some defects, as indicated by the Raman spectrum and a D peak fairly significant (Fig. 6). All the Raman spectra were obtained with a green laser of 532 nm. These spectra have been taken at arbitrary points of the samples after graphene CVD process with an acquisition time of 30 s. The ratio between D peak versus G peak intensities ($I_D/I_G$) evaluates the amount of defects, and the ratio between 2D peak versus G peak intensity ($I_{2D}/I_G$) normally presents the number of graphene layers[34,35]. The exact position of peaks should also be taken into account when extracting the total of the Raman spectrum information, because the peak position varies depending on the substrate (whether the graphene is on Cu or Si) which makes more difficult the theoretical fit. As we can see, D peak is remarkable because the graphene deposited on Si has wrinkles and is polycrystalline. In Fig. 7, graphene on Si/Cu is studied by means of Raman mapping with its ratio $I_{2D}/I_G$. This image shows the number of graphene layers depending on the position. The domains of graphene on polycrystalline Si reach a dimension of 50 $\mu m^2$, while the domain areas of graphene on polycrystalline Cu are greater than 1mm$^2$. If we compare this picture with the corresponding image obtained with the optical microscope, we note that there are less graphene layers on the borders of the Cu crystals.

Although related works have been published[14,15], we consider that our monolayer graphene is obtained from *instant* and *very-low pressure pulses* of gas. Han et al. succeeded in growing graphene dividing the growth process into a sequence of short time slots during which methane was introduced at a constant flux (as well as a mixture of Ar/H$_2$), but they needed around 100-300 of these cycles to obtain graphene. In the case of Puretzky et al. they used acetylene as a carbon source and Ni films as a catalyst, but in spite of using pulses in a CVD system too, the pressure they used cannot be considered as very low pressure as the used pressures of 18 Pa.

## IV. SUMMARY AND CONCLUSIONS

The present work demonstrates the agreement between the monolayer-formation-time ($\tau$) equation to grow graphene through an alternative very low-pressure Pulsed-CVD system developed for this purpose. Graphene was successfully grown on a thin film of sputtered Cu/Ni and on c-Si; reducing the deposition time to the order of 10 s and the methane partial pressure up to $10^{-4}$ Pa. The Raman analysis, SEM and EDS assessed the only presence of large-area graphene (up to $10^4$ $\mu m^2$) of one-two layers by showing the characteristic 2D band and a ratio $I_{2D}/I_G \geq 1$.

Further work is necessary to optimize the theoretical approach of the monolayer-formation time equation: the sticking coefficient needs to be evaluated; as well as the very low-pressure Pulsed-CVD method: the importance of the copper layer thickness, the optimal annealing conditions, and the removing of the Cu/Ni during annealing after the CVD process to grow graphene only on silicon or silicon dioxide. This is especially important in the application of lithographic processes and the possibility to produce graphene-based electronic devices.

## ACKNOWLEDGMENTS

This work was supported by *AGAUR* of Generalitat de Catalunya (project 2009GR00185) and by *MICINN* of Spanish Government (project MAT2010-20468). To the staff of the *CCiT-UB* for their technical support in